\documentclass[preprint,showpacs,preprintnumbers,amsmath,amssymb]{revtex4}

\usepackage{graphicx}% Include figure files
\usepackage{dcolumn}% Align table columns on decimal point
\usepackage{bm}% bold math
 
\begin{document}

\noindent

\preprint{}

\title{Objective uncertainty relation with classical background in a statistical model}% Force line breaks with \\

\author{Agung Budiyono} 
\email{agungby@yahoo.com}

\affiliation{Jalan Emas 772 Growong Lor RT 04 RW 02 Juwana, Pati, 59185 Jawa Tengah, Indonesia}

\date{\today}% It is always \today, today,
             %  but any date may be explicitly specified

\begin{abstract}    
We show within a statistical model of quantization reported in the previous work based on Hamilton-Jacobi theory with a random constraint that the statistics of fluctuations of the actual trajectories around the classical trajectories in velocity and position spaces satisfy a reciprocal uncertainty relation. The relation is objective (observation independent) and implies the standard quantum mechanical uncertainty relation.
\end{abstract} 
 
% PACS, the Physics and Astronomy 
                             % Classification Scheme.
%Use showkeys class option if keyword
                              %display desired
\maketitle  

\section{Introduction}

In the previous work we have developed a statistical model of quantization for non-relativistic system of spin-less particles \cite{AgungSMQ0}. We assumed that there exists some universal background fields interacting with the system (whose physical nature is not our present concern), resulting in the stochastic motion of the latter. We then assumed that the Hamilton-Jacobi theory for ensemble of trajectories has to be subjected to a random constraint. We showed that given a Lagrangian and a specific type of constraint uniquely determined by the Lagrangian, the effective dynamics of the ensemble of trajectories in configuration space is governed by a Schr\"odinger equation from which we read-off a unique quantum Hamiltonian. The Born's statistical interpretation of wave function is valid by construction.      

Further, unlike canonical quantization whose physical meaning behind the formal mathematical rules of replacement of c-number (classical number) by q-number (quantum number/Hermitian operator) is not transparent, the statistical model of quantization reported in Ref. \cite{AgungSMQ0} can be directly interpreted as {\it a specific statistical deviations from ensemble of classical trajectories parameterized by an unbiased non-vanishing random variable} $\lambda$ \cite{AgungSMQ1}. $\lambda$ is just the Lagrange multiplier that arises in the Hamilton-Jacobi theory with a random constraint. The prediction of canonical quantization with a unique ordering is reproduced if the distribution of $\lambda$ takes the form:
\begin{equation}
P(\lambda)=\frac{1}{2}\delta(\lambda+\hbar)+\frac{1}{2}\delta(\lambda-\hbar), 
\label{God's coin}
\end{equation}
characterized by the Planck constant. For more general distribution of $\lambda\neq 0$ that deviates slightly from Eq. (\ref{God's coin}) yet is still unbiased, $P(\lambda)=P(-\lambda)$, the model suggests testable possible small corrections to the statistical predictions of canonical quantization \cite{AgungSMQ2}. 

It is then instructive to study the statistics of the deviations from the classical trajectories. We shall show in the present paper a kinematical feature of the above statistical model that given a wave function, the average of the deviations of the actual trajectories from the corresponding classical trajectories in velocity and position spaces satisfy an uncertainty relation in a formally similar fashion as the standard quantum mechanical uncertainty relation \cite{QMUR}. The uncertainty relation to be presented is however objective referring to no measurement (observation independent), and furthermore implies the standard quantum mechanical uncertainty relation. 

\section{General formalism}

Let us denote the classical Lagrangian of the system as $\underline{L}(q,\dot{q})$, where $q$ is the configuration of the system and $\dot{q}\doteq dq/dt$, with $t$ is time, is the velocity. For simplicity, we will consider system with only one degree of freedom. Extension to many degrees of freedom is straightforward. In the statistical model of quantization, the momentum $\underline{p}(q,\dot{q})\doteq\partial\underline{L}/\partial\dot{q}$ and the phase of the wave function $S(q,\lambda;t)$, a stochastic real-valued function satisfying a modified Hamilton-Jacobi equation \cite{AgungSMQ0}, is related to each other as follows (see Eq. (14) of Ref. \cite{AgungSMQ0}) \cite{1/2}: 
\begin{equation} 
\underline{p}(q,\dot{q})=\partial_qS+\frac{\lambda}{2}\frac{\partial_q\Omega}{\Omega},
\label{fundamental equation 0} 
\end{equation}
where $\Omega=\Omega(q,\lambda;t)$ is the joint-probability density of the fluctuations of $q$ and $\lambda$ at time $t$ which is assumed to be even in $\lambda$, $\Omega(q,\lambda;t)=\Omega(q,-\lambda;t)$. For our purpose it is sufficient to consider the case of a single particle of mass $m$ subjected to external potential $V(q)$. The Lagrangian then takes the form $\underline{L}=m\dot{q}^2/2-V(q)$ so that  $\underline{p}=m\dot{q}$. Equation (\ref{fundamental equation 0}) thus becomes 
\begin{equation}
m\dot{q}(q,\lambda;t)=\partial_qS+\frac{\lambda}{2}\frac{\partial_q\Omega}{\Omega}.
\label{fundamental equation}
\end{equation}
The classical limit corresponds to the regime when the second term of the right hand side is ignorable or formally when  $|\lambda|\ll 1$ so that one regains the classical relation $m\dot{q}\approx\partial_qS$. $\delta_{\dot{q}}(q,\lambda)\doteq(\dot{q}-\partial_qS/m)^2=\frac{\lambda^2}{4m^2}(\partial_q\Omega/\Omega)^2=\delta_{\dot{q}}(q,-\lambda)$ can thus be interpreted to give the deviations from the classical mechanics in velocity space.  

Let us then consider $\Omega$ at an arbitrary snapshot of time. For notational simplicity, we shall thus suppress the time dependence $\Omega(q,\lambda)$. Then, from the normalization of $\Omega$, $\int dqd\lambda\Omega(q,\lambda)=1$, and the assumption that $\Omega(\pm\infty,\lambda)=0$ for arbitrary value of $\lambda$, one has 
\begin{eqnarray}
-1=-\int dqd\lambda\Omega=\int dqd\lambda (q-q_0)\partial_q\Omega\hspace{0mm}\nonumber\\
=\int dqd\lambda \{(q-q_0)\sqrt{\Omega}\}\Big\{\frac{\partial_q\Omega}{\sqrt{\Omega}}\Big\}, 
\end{eqnarray}
where $q_0$ is an arbitrary real number and the integration over spatial coordinate is taken from $q=-\infty$ to $q=\infty$. The Schwartz inequality then implies 
\begin{equation}
\int dqd\lambda (q-q_0)^2\Omega\times\int dqd\lambda\Big(\frac{\partial_q\Omega}{\Omega}\Big)^2\Omega\ge 1. 
\label{Schwartz inequality}
\end{equation} 
Substituting Eq. (\ref{fundamental equation}) into Eq. (\ref{Schwartz inequality}) one directly gets  
\begin{equation}
\int dqd\lambda (q-q_0)^2\Omega\times\int dqd\lambda\frac{4}{\lambda^2}(m\dot{q}-\partial_qS)^2\Omega\ge 1.  
\label{uncertainty relation 0}
\end{equation}
As shown in Ref. \cite{AgungSMQ0}, the results of canonical quantization is reproduced by the statistical model if $\Omega(q,\lambda)=\rho(q,|\lambda|)P(\lambda)$ where $P(\lambda)$ is given by Eq. (\ref{God's coin}) and $\rho(q,\hbar)$ is related to the quantum mechanical wave function $\Psi_Q(q,\hbar)$ satisfying a Schr\"odinger equation through the Born's statistics $\rho(q,\hbar)=|\Psi_Q(q,\hbar)|^2$ \cite{AgungSMQ0,AgungSMQ1,AgungSMQ2}. In this case, Eq. (\ref{uncertainty relation 0}) reduces into
\begin{equation}
\int dq (q-q_0)^2\rho(q,\hbar)\times\int dq(\dot{q}-\partial_qS_Q/m)^2\rho(q,\hbar)\ge\frac{\hbar^2}{4m^2},
\label{uncertainty relation 1}
\end{equation} 
where $S_Q(q;t)=S(q,\pm\hbar;t)$ is the quantum mechanical phase.    

If we take $q_0$ as the configuration of the corresponding classical system at the time of interest, then, as claimed, Eq. (\ref{uncertainty relation 1}) is just a reciprocal uncertainty relation between the average deviations of the actual trajectory from the corresponding classical trajectory in velocity and position spaces. Notice that we have only used Eq. (\ref{fundamental equation}) in deriving Eq. (\ref{uncertainty relation 1}). No dynamics is involved. The uncertainty relation thus directly reflects the kinematics of the statistical model of quantization. Further, given $S(q,\lambda)$, the permissible value of $\dot{q}(q,\lambda)$ is determined by the choice of $\Omega(q,\lambda)$ through Eq. (\ref{fundamental equation}). $S(q,\lambda)$ thus can be regarded as a parameter for an equivalent class of ensemble, or identically prepared ensemble, which determines the relation between $\Omega(q,\lambda)$ and $\dot{q}$. Keeping these in mind, one can thus interpret the uncertainty relation of Eq. (\ref{uncertainty relation 0}) as the impossibility to prepare an ensemble, using identical procedure defined by choosing $S(q,\lambda)$, that violates the relation.  

As is clear from the derivation, Eq. (\ref{uncertainty relation 1}) is valid for arbitrary choice of $q_0$. One can however show that when $q_0=\int dq q\rho$, $\int dq(q-q_0)^2\rho$ takes its minimum value: $\int dq (q-q_0)^2\rho\ge\int dq (q-\int dq q\rho)^2\rho$ \cite{lms}. This point is in particular relevant when we derive the usual quantum mechanical uncertainty relation from Eq. (\ref{uncertainty relation 1}). This will be done in Section IV.  

\section{Single slit experiment}

To illustrate the above interpretation, let us assume that the distribution of $q$ is given by a Gaussian as follows:
\begin{equation}
\Omega(q,\lambda)=\sqrt{\frac{a(|\lambda|)}{\pi}}\exp(-a(|\lambda|)q^2)P(\lambda), 
\label{Gaussian probability density}
\end{equation}  
where $a$ is a positive definite real-valued function of $|\lambda|$ and for simplicity we have assumed that the Gaussian is centered at the origin. It is evidently normalized $\int d\lambda dq\Omega=\int d\lambda P(\lambda)=1$. The variance of $q$ is thus given by $\Sigma_q(|\lambda|)=1/(2a(|\lambda|))$. One can show that when $a=m\omega/|\lambda|$ where $\omega$ is independent of $\lambda$, then $\Psi(q,\lambda)=\sqrt{\Omega}$ is the ground state of a harmonic oscillator which in the statistical model has a $\lambda$-parameterized quantum Hamiltonian $\hat{H}(|\lambda|)=-(\lambda^2/2m)\partial_q^2+m\omega^2q^2/2$ \cite{AgungSMQ0,AgungSMQ1,AgungSMQ2}. The standard quantum mechanical ground state wave function is reproduced in the case when $P(\lambda)$ in Eq. (\ref{Gaussian probability density}) is given by Eq. (\ref{God's coin}) so that one regains the standard quantum Hamiltonian for the harmonic oscillator $\hat{H}(\hbar)=-(\hbar^2/2m)\partial_q^2+m\omega^2q^2/2$ with frequency $\omega$. 

Inserting Eq. (\ref{Gaussian probability density}) into Eq. (\ref{fundamental equation}), one gets 
\begin{equation}
m\dot{q}=\partial_qS-\lambda aq.  
\end{equation}
For simplicity, let us then proceed to prepare an ensemble of system using identical procedure such that $S$ does not depend on $q$, namely $\partial_qS=0$. One thus has
\begin{equation}
\dot{q}=-a\lambda q/m.
\label{momentum for classically resting particle}
\end{equation}
Hence, the statistical model predicts that such identical preparation will unavoidably lead to an actual velocity field given by Eq. (\ref{momentum for classically resting particle}). The actual velocity is thus fluctuating around zero with vanishing average. Notice that the magnitude of the fluctuations is proportional to $a$ which is the inverse of the variance of the Gaussian distribution of position of Eq. (\ref{Gaussian probability density}). 
 
Let us then calculate the distribution of the actual velocity as a result of such identical preparation. Denoting the probability density that the velocity is $\dot{q}$ as $\widetilde{\rho}(\dot{q})$, using Eqs. (\ref{momentum for classically resting particle}) and (\ref{Gaussian probability density}), one has, up to a normalization constant, 
\begin{equation}
\widetilde{\rho}(\dot{q})\sim\int dqd\lambda\delta(\dot{q}-(-a\lambda q/m))\Omega(q,\lambda)\sim\int d\lambda P(\lambda)e^{-\frac{m^2\dot{q}^2}{a\lambda^2}}. 
\label{Gaussian momentum distribution}
\end{equation}
In particular, in the case when $P(\lambda)$ is given by Eq. (\ref{God's coin}), one has   
\begin{equation}
\widetilde{\rho}(\dot{q})\sim\exp\Big(-\frac{m^2\dot{q}^2}{a\hbar^2}\Big).
\label{Gaussian momentum distribution quantum} 
\end{equation}
The actual velocity of the particle is thus distributed according to a Gaussian with a variance $\Sigma_{\dot{q}}(\hbar)=a\hbar^2/(2m^2)$. One can see from Eqs. (\ref{Gaussian probability density}) and (\ref{Gaussian momentum distribution quantum}) that the variances of the fluctuations of position and velocity satisfy $\Sigma_q(\hbar)\Sigma_{\dot{q}}(\hbar)=\hbar^2/4m^2$. Hence, in this case, the equality in Eq. (\ref{uncertainty relation 1}) is achieved. 

The well-known single slit experiment gives an example of the above discussion. The experiment can be interpreted as a method to identically prepare (select/filter) an ensemble of particle characterized by $S$ which is independent of $q$, $\partial_qS=0$, and each has a definite position $q_0$. The former is obtained by preparing a beam of planar wave in the direction perpendicular to the plane of the slit. On the other hand, the latter is obtained by narrowing the width of the slit. As verified by experiment, the ensemble obtained by such identical preparation however is limited by the uncertainty relation of the type of Eq. (\ref{uncertainty relation 1}). Namely, selecting ensemble of trajectories so that the position of the particle is closer to the target position $q_0$ by narrowing the width of the slit, will automatically results in an ensemble with larger uncertainty of the actual velocity and vice versa. Let us remark that there is no measurement of position and velocity involved in the above experiment.    

\section{Discussion}

Let us compare the above derived uncertainty relation with the standard uncertainty relation of quantum theory. First, in the pragmatical interpretation, the latter refers to the statistics of results of measurement over an ensemble of identically prepared system. In the case of single slit experiment for example, one performs measurement  over the ensemble that is selected by the slit. Hence, one makes position measurement over half of the ensemble and momentum measurement over the other half, and calculate the statistical spread of the results, assuming that the ensemble is infinitely large, to get, by the virtue of the canonical uncertainty relation \cite{QMUR}
\begin{equation}
\Delta_q\Delta_p\ge\hbar^2/4, 
\label{QM uncertainty relation}
\end{equation}
where $\Delta_{q(p)}$ is the variance of the results of measurement of position (momentum). In this context, the above relation clearly has nothing to do with the limitation of simultaneous measurement of position and momentum, usually called as Heisenberg uncertainty principle \cite{HUP}. See also Refs. \cite{Isham book,Ozawa} for the discussion concerning this issue.  
 
While, as mentioned above, the operational interpretation of Eq. (\ref{QM uncertainty relation}) is clear, owing to the ambiguity of the physical interpretation of quantum mechanical wave function, there are several different physical interpretations of the relation of Eq. (\ref{QM uncertainty relation}). See for example the discussion in Ref. \cite{Isham book}. By contrast, by construction, the physical meaning of the uncertainty relation of Eq. (\ref{uncertainty relation 1}) is straightforward. It directly reflects the actual distribution of $q$ and $\dot{q}$ prior to the measurement. In this context, we say that the uncertainty relation of Eq. (\ref{uncertainty relation 1}) is objective (observation independent). Further, while the quantum mechanical uncertainty relation does not refer, at least directly, to a classical background, Eq. (\ref{uncertainty relation 1}) evidently refers to the fluctuations around the classical background (ensemble of classical trajectories). In this regards, we say that the relation is explicitly classical-context-dependent. 

Next, one can show that the uncertainty relation of Eq. (\ref{uncertainty relation 1}) gives the lower bound to the quantum mechanical uncertainty relation as follows. First, as shown in Ref. \cite{AgungSMQ0}, the statistical model of quantization reproduces the statistical prediction of quantum mechanics when $\Omega$ is factorizable  $\Omega(q,\lambda)=\rho(q,|\lambda|)P(\lambda)$ and $P(\lambda)$ is given by Eq. (\ref{God's coin}). The quantum mechanical wave function can then be written as $\Psi_Q=\sqrt{\rho}\exp(iS_Q/\hbar)$ where $S_Q=S(q,\pm\hbar)$ satisfies a modified Hamilton-Jacobi equation. One can then show straightforwardly the following mathematical identity: 
\begin{eqnarray}
\Delta_p=\langle(\hat{p}-\langle\hat{p}\rangle_Q)^2\rangle_Q\hspace{40mm}\nonumber\\
=\Big\langle\Big(\frac{\hbar}{2}\frac{\partial_q\rho}{\rho}\Big)^2\Big\rangle_S+\langle(\partial_qS_Q-\langle \partial_qS_Q\rangle_S)^2\rangle_S, 
\label{QF vs SM fluctuations 0} 
\end{eqnarray} 
where $\hat{p}\doteq-i\hbar\partial_q$ is the quantum mechanical momentum operator, $\langle\hat{\circ}\rangle_Q\doteq\langle\Psi_Q|\hat{\circ}|\Psi_Q\rangle$ is the quantum mechanical average, and $\langle\star\rangle_S\doteq\int dq\rho\star$ is the statistical average of $\star$ over $\rho$. Here we have used the identity $\langle\hat{p}\rangle_Q=\langle \partial_qS_Q\rangle_S$. For example, in the case of Gaussian wave function discussed before, one has $S_Q=0$ so that the second term on the right hand side of Eq. (\ref{QF vs SM fluctuations 0}) is vanishing. 
 
Taking into account Eq. (\ref{fundamental equation}) for the case $\Omega=\rho(q,|\lambda|)P(\lambda)$ where $P(\lambda)$ is given by Eq. (\ref{God's coin}), Eq. (\ref{QF vs SM fluctuations 0}) then becomes 
\begin{eqnarray}
\Delta_p=\langle(m\dot{q}-\partial_qS_Q)^2\rangle_S+\langle(\partial_qS_Q-\langle \partial_qS_Q\rangle_S)^2\rangle_S. 
\label{QF vs SM fluctuations} 
\end{eqnarray}
On the other hand, as discussed in Section II, taking $q_0=\langle q\rangle_S=\langle q\rangle_Q$, then $\langle (q-q_0)^2\rangle$ takes its minimum given by $\langle(q-q_0)^2\rangle_S=\langle(q-\langle q\rangle_Q)^2\rangle_Q=\Delta_q$. Keeping in mind this and the fact that the second term on the right hand side of Eq. (\ref{QF vs SM fluctuations}) is non-negative, one can see that the uncertainty relation of Eq. (\ref{uncertainty relation 1}) implies the standard quantum mechanical uncertainty relation: 
\begin{eqnarray}
\Delta_q\Delta_p=\langle(q-q_0)^2\rangle_S\Big(\langle(m\dot{q}-\partial_qS_Q)^2\rangle_S\nonumber\\
+\langle(\partial_qS_Q-\langle \partial_qS_Q\rangle_S)^2\rangle_S\Big)\nonumber\\
\ge \langle(q-q_0)^2\rangle_S\langle(m\dot{q}-\partial_qS_Q)^2\rangle_S\ge\hbar^2/4. 
\end{eqnarray} 

Note also that the first term of Eq. (\ref{QF vs SM fluctuations 0}) takes the form of the Fisher information for translations of $q$ with probability density $\rho(q;\hbar)$ multiplied by $\hbar^2/4$ \cite{Fisher inf.}. In the statistical model of quantization, the Fisher information is thus shown to be proportional to the average deviation of the actual trajectories from the corresponding classical trajectories in velocity space. In this context, the uncertainty relation of Eq. (\ref{uncertainty relation 1}) is formally just the Cramer-Rao inequality \cite{Cramer-Rao}. The relation between Fisher information and quantum fluctuations are also reported with different contexts in Refs. \cite{Frieden}.      

A formally similar relation as in Eq. (\ref{uncertainty relation 1}) is also obtained in Nelsonian stochastic mechanics \cite{Fenyes,Weizel,Kersaw,Nelson,de la Pena SM}. In this model of quantum fluctuations, first one assumes that the stochasticity implies non-differentiable Brownian trajectories. It is then impossible to define a conventional velocity of the Brownian particle. Instead, one then defines mean forward $v_+$ and mean backward $v_-$ velocities whose difference gives the so-called osmotic velocity 
\begin{equation}
u\doteq\frac{1}{2}(v_+-v_-)=\frac{\hbar}{2m}\frac{\partial_q\rho}{\rho}. 
\label{osmotic velocity}
\end{equation}
Further, in this Nelsonian stochastic mechanics, the current velocity, which corresponds to the effective velocity of the statistical model of Ref. \cite{AgungSMQ0}, is defined as $v\doteq(v_++v_-)/2$. If the trajectory is smooth (differentiable), then $v_+=v_-$ so that the osmotic velocity is vanishing $u=0$. It is then straightforward to develop from Eq. (\ref{osmotic velocity}) an uncertainty relation \cite{de la Pena SMUR,Falco SMUR,Golin SMUR,Allah SMUR}
\begin{equation}
\langle (q-\langle q\rangle_S)^2\rangle_S\langle u^2\rangle_S\ge\frac{\hbar^2}{4m^2}. 
\label{stochastic uncertainty relation} 
\end{equation}
Note that Eq. (\ref{osmotic velocity}) implies $\langle u\rangle_S=0$. 

The above uncertainty relation thus arises due to stochasticity of the dynamics which in particular implies the absence of regular trajectory. The latter leads to the necessity to have forward and backward diffusive stochastic processes and naturally gives the definition of the osmotic velocity of Eq. (\ref{osmotic velocity}). By contrast, while the uncertainty relation developed in the present paper is caused by the presence of a random constraint \cite{AgungSMQ0}, we assume that the trajectory is as smooth as in classical mechanics which allows us to have the usual definition of velocity. Let us note however that $\langle u^2\rangle_S$ in Eq. (\ref{stochastic uncertainty relation}) corresponds to $\langle (\dot{q}-\partial_qS_Q/m)^2\rangle_S$ of the present statistical model, measuring the deviations of the ensemble of actual trajectories from the classical trajectories in velocity space. 

Next, we have shown in Refs. \cite{AgungSMQ0} that in the special case when $\Omega$ is factorizable as $\Omega(q,\lambda)=\rho(q,|\lambda|)P(\lambda)$ and $P(\lambda)$ is given by Eq. (\ref{God's coin}), the statistical model is effectively equivalent to the pilot-wave model \cite{pilot-wave theory}. Namely, the {\it effective} velocity of the particle defined as $(\dot{q}(\hbar)+\dot{q}(-\hbar))/2$ in the statistical model is numerically equal to the {\it actual} velocity of the particle in pilot-wave theory. This intimate relationship is further reflected by the fact implied by Eq. (\ref{fundamental equation}) that the average deviations from the ensemble of classical trajectories in velocity space within the statistical model can be rewritten as follows:
\begin{eqnarray}
\frac{1}{2m}\langle(m\dot{q}-\partial_qS_Q)^2\rangle_S=\frac{1}{2m}\Big\langle\Big(\frac{\lambda}{2}\frac{\partial_q\rho}{\rho}\Big)^2\Big\rangle_S=\langle U\rangle_S,\nonumber\\
\mbox{where}\hspace{2mm} U\doteq -\frac{\hbar^2}{2m}\frac{\partial_q^2\sqrt{\rho}}{\sqrt{\rho}}. \hspace{20mm} 
\label{flucuations of momentum and quantum potential} 
\end{eqnarray}
Here $U$ is the so-called quantum potential which in the pilot-wave theory is argued to be responsible for all peculiar quantum phenomena. It is remarkable that the deviation from the ensemble of classical trajectory in velocity space is measured by the average of quantum potential. Hence, in both of the present statistical model and pilot-wave model, the classical limit is obtained when the quantum potential is vanishing.   

\section{Conclusion}

We have thus developed, within the statistical model of quantization reported in Ref. \cite{AgungSMQ0}, an uncertainty relation which is objective and implies the quantum mechanical uncertainty relation. There is no notion of non-commutativity between pair of so-called quantum observables as in standard formalism of quantum theory \cite{Isham book}, nor there is a need to assume forward and backward diffusion processes as in the Nelsonian stochastic mechanics \cite{de la Pena SMUR,Falco SMUR,Golin SMUR,Allah SMUR}. Bohr complementarity is argued to apply not only to describe the statistics of results of measurement, but is extended to the distribution of the {\it actual} position and velocity. Further, the uncertainty relation is classically contextual in the sense that it describes the fluctuations around the ensemble of classical trajectories. Given a quantum mechanical state (wave function), it therefore provides an explicit measure to the degree of imprecision of classical mechanics or equivalently the degree of quantum-ness of the state. It is then interesting to see in the future work the implication of assuming $P(\lambda)$ which deviates from Eq. (\ref{God's coin}).       

\begin{acknowledgments}  

The present research was initiated when the author held an appointment with RIKEN under the FPR program. 

\end{acknowledgments}


\begin{thebibliography}{10}
\bibitem{1/2} In Ref. \cite{AgungSMQ0} we have used a scale so that $\lambda$ in the present paper is related to $\lambda$ in that paper as $\lambda\mapsto 2\lambda$.
\end{thebibliography}
\end{document}